\documentclass[prd,reprint,aps,amsmath,amssymb,showpacs,nofootinbib,preprintnumbers,superscriptaddress]{revtex4-1}
\usepackage[utf8x]{inputenc}
\usepackage{graphicx}
\usepackage{enumitem}
\usepackage{slashed}
\usepackage{url}
\usepackage[pdftex]{hyperref}
\usepackage[T1]{fontenc}
\usepackage{bm}
\usepackage{color}
\usepackage{bbm}

\newcommand{\fslash}[1]{\!\not\!{#1}}

\begin{document}

\title{Neutron stars exclude light dark baryons}
\preprint{PITT-PACC-1803}
\preprint{INT-PUB-18-007}

\author{David McKeen}
\email{dmckeen@pitt.edu}
\affiliation{Pittsburgh Particle Physics, Astrophysics, and Cosmology Center, Department of Physics and Astronomy, University of Pittsburgh, PA 15260, USA}

\author{Ann E. Nelson}
\email{aenelson@uw.edu}
\affiliation{Department of Physics, University of Washington, Seattle, WA 98195, USA}

\author{Sanjay Reddy}
\email{sareddy@uw.edu}
\affiliation{Institute for Nuclear Theory, University of Washington, Seattle, WA 98195, USA}

\author{Dake Zhou}
\email{zdk@uw.edu}
\affiliation{Department of Physics, University of Washington, Seattle, WA 98195, USA}
\affiliation{Institute for Nuclear Theory, University of Washington, Seattle, WA 98195, USA}
\date{\today}

\begin{abstract}
Exotic new particles carrying baryon number and with mass of order the nucleon mass have been proposed for various reasons including baryogenesis,  dark matter,  mirror worlds, and the neutron lifetime puzzle.
We show that the existence of neutron stars with mass greater than 0.7 $M_\odot$  places severe constraints on such   particles, requiring them to be heavier than 1.2 GeV or to have strongly repulsive self-interactions.
\end{abstract}

%\pacs{}
\maketitle

\section{Introduction}
Exotic states that carry baryon number and have masses below a few GeV have been theorized in a number of contexts, such as asymmetric dark matter~\cite{Shelton:2010ta,*Zurek:2013wia,Davoudiasl:2010am,*Davoudiasl:2011fj}, mirror worlds~\cite{Berezhiani:2005hv}, neutron-antineutron oscillations~\cite{McKeen:2015cuz} or in nucleon decays~\cite{Davoudiasl:2014gfa}.  In general, such states are highly constrained because they can drastically alter the properties of normal baryonic matter, in particular, if too light,  they can potentially render normal matter unstable. We currently understand that matter is observationally stable because the standard model (accidentally) conserves baryon number--this ensures that the proton, the lightest baryon, does not decay (up to effects caused by higher dimensional operators that violate baryon number).

Now, consider the simple case of a single new fermion state, $\chi$, that is electrically neutral, carries unit baryon number, and carries no other conserved charge. (Note that a new boson carrying baryon number does not lead to proton decay as long as lepton number is conserved.) Assuming that its couplings to ordinary matter are not highly suppressed, because of the conservation of baryon number and electric charge, it must have a mass larger than the difference between the proton and electron masses, $m_\chi>m_p-m_e=937.76~\rm MeV$, in order to not destabilize the proton. In fact, a slightly stronger lower bound on $m_\chi$ comes from the stability of the weakly bound $^9$Be nucleus: $m_\chi>937.90~\rm MeV$. If the $\chi$ mass is less than that of the neutron, $m_n=939.57~\rm MeV$, a new neutron decay channel can open up, $n\to\chi+\dots$, where the ellipsis includes other particles that allow the reaction to conserve (linear and angular) momentum.

It is interesting to note that if $m_\chi<m_p+m_e=938.78~\rm MeV$, $\chi$ is itself kept stable by the conservation of baryon number and electric charge. It could therefore be a potential candidate for the dark matter, which we know to be electrically neutral and stable on the timescale of the age of the Universe. It is compelling that in such a situation that the stability of normal matter and of dark matter is ensured by the {\it same} symmetry: baryon number.

The potential existence of a new decay channel for the neutron has recently received attention as a solution to the $4\sigma$ discrepancy between values of the neutron lifetime measured using two different techniques, the ``bottle'' and ``beam'' methods~\cite{Berezhiani:2005hv,Serebrov:2007gw,Fornal:2018eol}. The ``bottle'' method, which counts the number of neutrons that remain in a trap as a function of time and is therefore sensitive to the total neutron width gives $\tau_n^{\rm bottle}=879.6\pm0.6~\rm s$~\cite{Mampe:1993an,*Serebrov:2004zf,*Pichlmaier:2010zz,*Steyerl:2012zz,*Arzumanov:2015tea}. The ``beam'' method counts the rate of protons emitted in a fixed volume by a beam of neutrons, thus measuring only the $\beta$-decay rate of the neutron, results in $\tau_n^{\rm beam}=888.0\pm2.0~\rm s$~\cite{Byrne:1990ig,*Yue:2013qrc,*Byrne:1996zz}. These two measurements can be reconciled by postulating a new decay mode for the neutron, such as $n\to\chi+\dots$, with a branching fraction
\begin{equation}
{\rm Br}_{n\to\chi}=1-\frac{\tau_n^{\rm bottle}}{\tau_n^{\rm beam}}=\left(0.9\pm0.2\right)\times10^{-2}.
\end{equation}
However, a recent reevaluation of the prediction for the neutron lifetime from post 2002 measurements of the neutron $g_A$ concludes that any nonstandard branching for the neutron is limited to less than $2.7\times10^{-3}$ at 95\% CL~\cite{Czarnecki:2018okw}.

In this work we note that  a new state that carries baryon number and has a mass close to the neutron's can drastically affect the properties of nuclear matter at densities seen in the interiors of neutron stars. In neutron stars the neutron chemical potential can be significantly larger than $m_n$, reaching values $ \simeq 2 $ GeV in the heaviest neutron stars \cite{LattimerPrakash:2010}.  Thus any exotic particle that carries baryon number and has a mass $\lesssim 2 $ GeV  will have a large abundance if in chemical equilibrium. Because they replace neutrons, their presence will soften the equation of state of dense matter by reducing the neutron Fermi energy and pressure, while contributing to an increase in the energy density. This will in turn reduce the maximum mass of neutron stars from those obtained using  standard equations of state for nuclear matter. As we shall show below, even a modest reduction in the pressure at high density can dramatically lower the maximum mass to a value that is significantly smaller than the observed heaviest neutron stars with  masses $\simeq 2~M_\odot$ \cite{Demorest:2010bx,Antoniadis:2013pzd}.

The remainder of this paper is organized as follows. In section \S~\ref{Model} we describe a simple model of fermion dark matter which is charged under baryon number. In section \S~\ref{nstar} we show the results of a computation of the effects of such a fermion on mass radius relation and maximum mass of neutron stars. Possible extensions of these constraints, future work, and ways to avoid the constraints are  described in the conclusions,  \S~\ref{conclude}.

\section{Model}
\label{Model}
We begin by considering a simple model with a single neutral Dirac fermion, $\chi$, that carries unit baryon number. As mentioned above, $m_\chi>937.90~\rm MeV$ so as to not destabilize $^9$Be. The relevant terms in the effective Lagrangian involving the neutron are
\begin{equation}
{\cal L}_{\rm eff}=\bar n\left(i\!\fslash{\partial}-m_n\right)n+\bar\chi\left(i\!\fslash{\partial}-m_\chi\right)\chi-\delta\left(\bar\chi n+\bar n\chi\right),
\end{equation}
where $\delta$ is a coupling determined by the underlying theory. A simple UV completion~\cite{Arnold:2012sd,McKeen:2015cuz} of this involves integrating out a scalar diquark   coupled to $u$ and $d$ quarks as well as to $d$ and $\chi$, generating the four fermion operator
\begin{equation}
\frac{1}{\Lambda^2}\bar\chi udd.
\label{eq:4fermi}
\end{equation}
Matching this onto the effective theory gives
\begin{equation}
\delta\sim \frac{0.01~{\rm GeV}^3}{\Lambda^2}.
\end{equation}
In what follows, we assume that this coupling between $n$ and $\chi$ is small, in particular $|\delta|\ll |\Delta m|$, where $\Delta m\equiv m_n-m_\chi$. This coupling leads to a mixing between $n$ and $\chi$ and the mass terms are diagonalized by taking $n\to n+\theta\chi$, $\chi\to\chi-\theta n$, where the mixing angle is $\theta=\delta/\Delta m$.

If $m_\chi<m_n$, a new decay mode for the neutron opens up, $n\to\chi\gamma$. In addition, if $m_\chi<m_p+m_e=938.78~\rm MeV$, $\chi$ is stable. The new decay mode for the neutron comes from the neutron magnetic dipole moment operator, which, after the mass matrix is diagonalized contains the term
\begin{equation}
\mu_n\theta\bar\chi\sigma^{\mu\nu}nF_{\mu\nu},
\end{equation}
where $\mu_n=-1.91e/(2m_p)=-0.31~{\rm GeV}^{-1}$ is the neutron magnetic dipole moment. The partial width for $n\to\chi\gamma$ is
\begin{equation}
\Gamma_{n\to\chi\gamma}=\frac{\mu_n^2\theta^2m_n^3}{16\pi}\left(1-\frac{m_\chi^2}{m_n^2}\right)^3\simeq\frac{\mu_n^2\theta^2\Delta m^3}{2\pi}.
\label{eq:GammaFree}
\end{equation}
Given a total width of $\Gamma_n=1/\tau_n^{\rm bottle}=(879.6~{\rm s})^{-1}$, the branching ratio for the neutron to decay into $\chi\gamma$ is
\begin{equation}
\begin{aligned}
{\rm Br}_{n\to\chi\gamma}&=0.01\left(\frac{\Delta m}{1~\rm MeV}\right)^3\left(\frac{\theta}{7\times10^{-10}}\right)^2
\\
&=0.01\left(\frac{\Delta m}{1~\rm MeV}\right)\left(\frac{\delta}{7\times10^{-13}~\rm GeV}\right)^2.
\end{aligned}
\end{equation}

Thus, we see that for $m_n-m_\chi\sim 1~\rm MeV$, a mixing angle of order $10^{-9}$, or a $n$-$\chi$ coupling of about $10^{-12}~\rm GeV$ can explain the neutron lifetime anomaly.\footnote{We note here why a model with Dirac $\chi$ where baryon number is conserved is necessary. If instead $\chi$ where Majorana with $\theta=10^{-9}$ and $\Delta m=1~\rm MeV$, a $\Delta B=2$ $n$-$\bar n$ transition amplitude of roughly $\theta^2\Delta m\sim 10^{-21}~\rm GeV$ would arise. This is many orders of magnitude larger than the experimental upper bound of $10^{-33}~\rm GeV$.} This value of $\delta$ corresponds to a scale for the four fermion interaction of Eq.~(\ref{eq:4fermi}) of $\Lambda\sim10^5~\rm GeV$. We note here, however, that a very recent search for the decay $n\to\chi\gamma$ using ultracold neutrons sets a limit on this branching, for $937.90~{\rm MeV}<m_\chi<938.78~\rm MeV$, of roughly $10^{-3}$~\cite{Tang:2018eln}.

Although $\delta\sim10^{-12}~\rm GeV$ is a small coupling between the neutron and $\chi$, it can lead to the efficient conversion of neutrons into $\chi$'s in the high density environments encountered inside neutron stars. In addition, because of the large neutron chemical potential inside neutron stars, the conversion $n\to\chi$ can take place there even for $m_\chi>m_n$ where free neutron decays are kinematically blocked.

We investigate the effects of a $\chi$-$n$ coupling on neutron stars in the next section.

\section{Neutron stars}
\label{nstar}
The structure of neutron stars is determined by the equation of state (EOS)  of dense matter which specifies the relationship between pressure $P$ and energy density $\epsilon$. For a given EOS,  $P(\epsilon)$, the Tolman Oppenheimer Volkoff (TOV) equations of general relativistic hydrostatic structure can be be solved numerically to obtain the mass-radius curves~\cite{Oppenheimer:1939ne,Tolman:1939jz}. Although there remain large uncertainties associated with the EOS at supranuclear density, the EOS up to nuclear saturation density $n_s \simeq 0.16$ fm$^{-3}$ can be calculated using nuclear Hamiltonians and non-relativistic quantum many-body theory to obtain $P_{\rm nuc}(\epsilon_{\rm nuc})$~\cite{Akmal:1998cf,Gandolfi:2013baa,Hebeler:2013nza}. Further, absent phase transitions to new states of matter, modern nuclear EOSs are able to estimate uncertainties associated with the extrapolation to high density since they account for two and three body nuclear forces consistently, and are based on a systematic operator expansion rooted in effective field theory~\cite{Weinberg:1991um}. In what follows we shall use a modern EOS and demonstrate that, despite the uncertainty at supra-nuclear density, the observation of massive neutron stars with $M_{NS}\simeq 2 ~M_\odot$   rules out the existence of a weakly interacting dark matter candidate which carries baryon number and has a mass in the range $937.90~{\rm MeV}<m_\chi<938.78~\rm MeV$. In fact, we shall find that any such weakly interacting particle with mass $m_\chi \lesssim 1.2 $  GeV can be excluded.

\begin{figure}
\includegraphics[width=\linewidth]{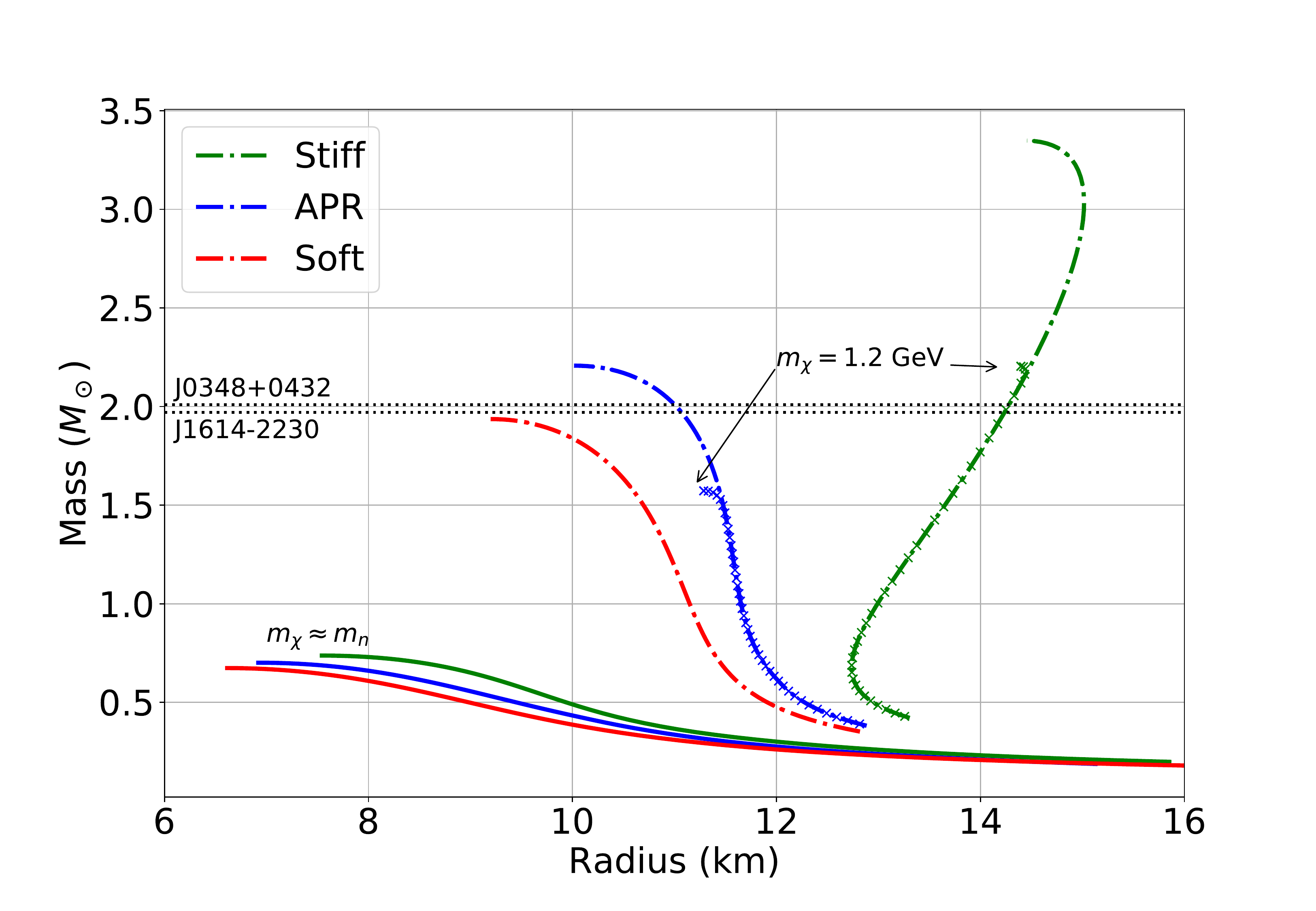}
\caption{The mass-radius relationship for selected nuclear EOS and resulting hybrid configurations. The standard nuclear matter relationships are shown as dash-dotted curves. The ``Stiff'' EOS makes a second order transition to a causal EoS at $n_B=1.5n_s$. This is the stiffest possible EOS and predicts a maximum mass $\simeq 3.3~M_\odot$. Adding a dark baryon with $m_\chi=938~\rm MeV$ results in the solid curves, which differ by their nuclear EOS. Even for the extremely stiff EOS, the maximum mass of hybrid stars containing non-interacting dark neutrons does not exceed $0.8M_\odot$. The measured masses of the two most massive neutron stars J0348+0432 and J1614-2230 are also shown.}\label{fig:MR}
\end{figure}

In Fig.~\ref{fig:MR} we show the mass-radius curve for neutron stars predicted by the standard nuclear EOS as dash-dotted curves. The curve labelled APR was obtained with a widely used nuclear EOS described in Ref.~\cite{Akmal:1998cf}. The curves labelled ``Soft'' and ``Stiff'' are the extreme possibilities consistent with our current understanding of uncertainties associated with the nuclear interactions up to $1.5 n_s$. The curves terminate at the maximum mass. For the softest nuclear equation of state just falls short of making a $2 ~M_\odot$ neutron star. The curve labelled ``Stiff'' is obtained by using the nuclear EOS that produces that largest pressure up to  $1.5 n_s$, and at higher density we use the maximally stiff EOS with $P (\epsilon)=p_0+ (\epsilon-\epsilon_0)$ where $p_0$ and $\epsilon_0$ are the pressure and energy density predicted by the nuclear EOS at $1.5 n_s$. For the maximally stiff EOS the speed of sound in the high density region $c_s = c$, and this construction produces the largest maximum mass of neutron stars compatible with nuclear physics. 

Any exotic neutron decay channel $ n \rightarrow \chi + \cdots$ which makes even a small contribution to the neutron width, of order the inverse lifetime of a neutron star,  will be fast enough to ensure that $\chi$ is equilibrium inside the star. The typical age $t_{NS}$ of old observed neutron stars    is  $t_{\rm NS}\approx 10^6-10^8$ years.  In a dense medium, due to strong interactions, the dispersion relation of the neutron can be written as  $\omega_n(p)=\sqrt{p^2+m_n^2} + \Sigma_r + i \Sigma_i$ where $\Sigma_r$ and $\Sigma_i$ are the real and  imaginary parts of its self-energy. The mixing angle is suppressed at finite density and is given by
\begin{equation}
\tilde{\theta}= \frac{\delta}{\sqrt{\widetilde{\Delta m}^2+\Sigma^2_i}}\,,
\end{equation}
where $\widetilde{ \Delta m} = \Delta m + \Sigma_r$. Since $\Sigma_r$ and $\Sigma_i$ are expected to be of the order of $10-100$ MeV at the densities attained inside neutron stars \cite{glendenning2000compact}, it is reasonable to expect the ratio $\tilde{\theta}/\theta$ to be in the range $0.01-0.1$. The rate of production of $\chi's$ in the neutron star interior due to neutron decay, defined in Eq.~\ref{eq:GammaFree}, is suppressed by the factor $(\tilde{\theta}/\theta)^2$ but enhanced by $(\widetilde{ \Delta m}/\Delta m)^3$ when $\widetilde{\Delta m}>\Delta m$. For $\widetilde{\Delta m} \approx 10$ MeV neutrons decay lifetime is  $<10^8$ yrs when  $\delta > 10^{-19}$ GeV, and it is safe to assume that for the phenomenologically interesting values of $\delta \simeq  10^{-14}-10^{-12}$ GeV, $\chi$ will come into equilibrium on a timescale $t \ll t_{\rm NS}$.\footnote{We delegate to future work a detailed calculation of the production rate for such small values of $\delta$ which may be interesting in other contexts.}

Because $\chi$ carries baryon number, in equilibrium its chemical potential $\mu_\chi = \mu_B$, where  $\mu_B$ is the baryon chemical potential. Given a nuclear EOS the baryon chemical potential is obtained using the thermodynamic relation  $\mu_B=(P_{\rm nuc}+\epsilon_{\rm nuc})/n_B$ where $n_B$ is the baryon number density. If $\chi$ is a Dirac fermion with spin $1/2$ and its interactions are weak, its Fermi momentum and energy density are given by
\begin{align}
k_{\rm F \chi}&=\sqrt{\mu_B^2-m^2_\chi}\,, \\
\epsilon_\chi&=\frac{1}{\pi^2}\int^{k_{\rm F \chi}}_0~dk~k^2\sqrt{k^2+m^2_\chi}\,, 
\end{align}
respectively. The dark neutron number density $n_\chi=k^3_{\rm F \chi}/3\pi^2$ and its pressure $P_\chi=-\epsilon_\chi+\mu_B n_\chi$.  The total pressure $P_{\rm tot}= P_{\rm nuc}+P_\chi$ and energy density $\epsilon_{\rm tot}=\epsilon_{\rm nuc}+\epsilon^{\rm kin}_\chi$ are easily obtained, and the TOV equations are solved again to determine the mass-radius relation for the hybrid stars containing an admixture of $\chi$ particles. The net result is a softer EOS where the pressure is lowered at a given a energy density because $\chi$ replaces neutrons and reduces their Fermi momentum and pressure. The results for $m_\chi=938$ MeV are shown in Fig.~\ref{fig:MR} as solid curves where the curves terminate at the maximum mass. We  allow the nuclear EOS to vary from maximally stiff to soft, and also show the results for the APR EOS. The striking feature is the large reduction in the maximum mass. This reduction is quite insensitive to the nuclear EOS. Even for the maximally stiff EOS, the presence of non-interacting dark neutrons reduce the maximum mass to values well below observed neutron star masses. Thus, a dark neutron with a $m_\chi \simeq 938$ MeV and weak interactions is robustly excluded.  For larger $m_\chi$ we can still obtain useful bounds as long as $m_\chi$ is smaller than the baryon chemical potential attained in the core.  For $m_\chi=1.2$ GeV we find as expected that the appearance of dark neutrons to delayed to supra-nuclear density, but as soon as they appear they destabilize the star. This is clearly seen by the behavior of the mass-radius relation labelled $m_\chi=1.2$ GeV and denoted by points represented as crosses. For the APR EOS the maximum mass is about  $1.6~M_\odot$ and for the maximally stiff EOS is is about  $2.2~M_\odot$.

Although interactions between $\chi$'s  and nucleons are necessarily weak,\footnote{In the model leading to the four fermion interaction of Eq.~(\ref{eq:4fermi}), $\chi$-nucleon interactions come from the  four fermion interaction $\bar\chi\chi\bar dd/\Lambda^2$. With $\Lambda\gtrsim 10^5~{\rm GeV}$ for $\delta\lesssim10^{-12}~{\rm GeV}$, this interaction is highly suppressed compared to nuclear strength interactions.} interactions between $\chi$'s could be strong. If $\chi$ is charged under a U(1) with coupling strength $g$ to a new gauge boson a mass $m_V$, repulsion between  between $\chi$'s modifies the EOS. In the mean field approximation, both the pressure and energy density are increased by
\begin{equation}
\Delta P_\chi=\Delta \epsilon_\chi= \frac{1}{2}\frac{g^2}{m^2_V}n^2_\chi\,,
\end{equation}
For strong coupling with $g\simeq 1$,  and small $m_V$ when the Compton wavelength of the gauge boson gets larger than the inter-particle distance this interaction energy will dominate. Under these conditions, the number density of $n_\chi \approx m^2_V \mu_B/g^2$ in equilibrium will be greatly reduced, and correspondingly its impact on the dense matter EOS will be negligible. 
Another possibility is that dark neutrons have interactions that mimic interactions between ordinary neutrons. In such a mirror scenario, we find that the maximum mass of neutron stars is $1.6 ~M_\odot$ for the APR EOS and $2.4 ~M_\odot$ for the maximally stiff EOS construction.

\section{Conclusions}
\label{conclude}
States that carry baryon number and have a mass close to the nucleons have been studied in several scenarios.   The extreme environments encountered in the interiors of neutron stars can readily produce such states. However, because these new states do not in general have the same interactions that neutrons do, they can lead to radically different EOS in neutron stars. In particular, new states will reduce the maximum possible neutron star mass which is consistent with a given nuclear EOS.

Simple scenarios where the dark baryons have a mass similar to that of the nucleon and are not charged under a new force do not allow for neutron stars with mass above $\sim 0.7-0.8 M_\odot$. This bound is in stark conflict with observation. Charging such dark baryons under a new force with a very light gauge mediator will result in interactions much larger than standard nuclear interactions and can greatly suppress their presence in dense matter.  This can mitigate their effect on the EOS enough to allow for neutron stars  as heavy as have been observed,  $\sim 2 M_\odot$. However, if such a new force is similar to nuclear forces as expected in a ``mirror world'' set up where the dark neutron has  the same self interactions as does the visible neutron, the maximum mass is still significantly reduced and one requires a very stiff high density EOS to produce $2 M_\odot$ neutron stars.   Interestingly, in the case where the dark baryons are stable dark matter, with $m_\chi\simeq 938~\rm MeV$,   nuclear strength self-interactions have been implicated to explain DM small scale structure puzzles (see, e.g.,~\cite{Tulin:2017ara} and references therein). 

Extensions of this work can easily be shown to constrain other possible new weakly interacting particles.  For instance in the ``hylogenesis'' baryogenesis scenario \cite{Davoudiasl:2010am,*Davoudiasl:2011fj}  there are two kinds of baryon number carrying dark matter particles, called  ``$Y$'' and ``$\Phi$'', which also carry another conserved charge for stability, but which have an allowed reaction $n+\gamma\leftrightarrow Y+\Phi$ .  Stability of matter places  a lower bound  of $937.90~\rm MeV$ on $m_Y+m_\Phi$.  The existence of observed neutron stars will place  a more stringent bound on $m_Y+m_\Phi$, which will be similar to the lower bound of 1.2 GeV we found on $m_\chi$.   Another type of new particle which would be constrained  would be a new  weak interacting neutral integer spin boson, called ``$\xi$'', with baryon number 1 and  interactions with ordinary matter which are not highly suppressed. As long as lepton number is conserved,  both $\xi$ and the proton are stable.  The stability of nuclei with atomic number A and charge Z  against decays of type $(A,Z)\rightarrow (A-2,Z) + 2\xi$ will place a lower bound of order the nucleon mass on $m_\xi$. Neutron stars, however, will  constrain $\xi$ to be heavier than the minimum chemical potential for neutrons in a 2 solar mass neutron star, or else neutrons  could convert to  $\xi$ particles and destabilize the star.

As noted earlier avoidance of such constraints is possible if the dark matter carries sufficiently repulsive self interactions. 
If the self repulsion of dark matter is large enough most of the mass of the star will remain in the form of neutrons and the effect of dark matter on the maximum mass will be small.

\begin{acknowledgments}
We thank Brian Batell, Pavel Fileviez Perez, and Maxim Pospelov for helpful discussions. We thank the Institute of Nuclear Theory and the organizers of the INT program {\it Neutron-Antineutron Oscillations: Appearance, Disappearance, and Baryogenesis} where this project began. DM is supported by PITT PACC through the Samuel P. Langley Fellowship. The work of AN was supported in part by the U.S. Department of Energy under Grant No. DE-SC0011637 and by the Kenneth K. Young Chair. SR and DZ are supported by the U.S. Department of Energy under Grant No. DE-FG02- 00ER41132. 
\end{acknowledgments}

\bibliography{darkneutrons}

\end{document}